\documentclass[prb,american,aps,twocolumn,showpacs,unsortedaddress]{revtex4}
\usepackage{graphics,bm}
\usepackage{amssymb}
\usepackage{epsfig}
\usepackage{epsfig}
\usepackage{epsf}
\usepackage{graphicx}
\usepackage{amsmath}
\usepackage{mathrsfs}
\usepackage{amsfonts}
\makeatletter

\newcommand{\beqa}{\begin{eqnarray}}
\newcommand{\eeqa}{\end{eqnarray}}
\newcommand{\beq}{\begin{equation}}
\newcommand{\eeq}{\end{equation}}

\begin{document}

\title{\bf{Quantum theory of spin waves in finite chiral spin chains}}

\author{A. Rold\'an-Molina}
\affiliation{Instituto de F\'isica, Pontificia Universidad Cat\'olica de Valpara\'iso, Avenida Brasil 2950, Casilla 4059, Valpara\'iso, Chile}
\author{M. J.  Santander}
\author{\'A.  S. N\'u\~{n}ez} 
\affiliation{Departamento de F\'isica, Facultad de Ciencias F\'isicas y Matem\'aticas, Universidad de Chile, Casilla 487-3, Santiago, Chile}
\author{J.  Fern\'andez-Rossier\footnote{Permanent Address: Departamento de F\'isica Aplicada, Universidad de Alicante}} 
\affiliation{International Iberian Nanotechnology Laboratory, Av. Mestre Jose Veiga, 4715-310 Braga, Portugal}

\begin{abstract}
 We calculate the effect of spin waves on the properties of finite size  spin chains with a chiral spin ground state  observed on bi-atomic Fe chains  deposited on Iridium(001).
The system is described with a  Heisenberg model supplemented with a Dzyaloshinskii-Moriya (DM) coupling and a uniaxial single ion anisotropy that presents a chiral spin ground state.  Spin waves are studied  using the Holstein-Primakoff (HP) boson representation of spin operators. Both the renormalized ground state and the elementary excitations are found by means of  Bogoliubov  transformation, as a function of the two variables that can be controlled experimentally,  the applied magnetic field and the chain length.
Three main results are found. 
First, because of the   non-collinear nature of the classical ground state, there is a {\bf significant}
zero point reduction of the  ground state magnetization of the spin spiral. 
Second, the two lowest energy spin waves are edge modes in the spin spiral state that, 
 above a critical field the results into a collinear ferromagnetic ground state, become confined bulk modes. Third, in the spin spiral state,  the spin wave spectrum exhibits  oscillatory behavior as function of the chain length with the same period of the spin helix. 
\end{abstract}

 \pacs{74.50.+r,03.75.Lm,75.30.Ds}

\maketitle
\section{Introduction} 

Because of the possibility of engineering  and probing spin chains, atom by atom,  using scanning tunneling microscope,
\cite{Hirjibehedin_Lutz_Science_2006,Serrate_Ferriani_natnano_2010,Khajetoorians_Wiebe_science_2011,wiesendanger,Khajetoorians_Wiebe_natphys_2012,Loth_Baumann_science_2012}  the study of spin chains is not only a crucial branch in the study  strong correlations and quantum magnetism\cite{Yosida,Auerbach}, but also a frontier in the research of atomic scale spintronics.\cite{JFR-NAT-MAT2013} 
Spin chains display a vast array of different magnetic states depending on the interplay between spin interactions, size of the chain and their dissipative coupling to the environment.  Thus,  experiments reveal that different spin chains can behave like quantum antiferromagnets,\cite{Hirjibehedin_Lutz_Science_2006} classical antiferromagnets,\cite{Khajetoorians_Wiebe_science_2011,Loth_Baumann_science_2012} and classical spin spirals.\cite{wiesendanger}  When quantum fluctuations do not quench the atomic magnetic moment,  classical information can be stored and manipulated in atomically engineered spin chains. Thus,  classical N\'eel states can be used  to store a bit of information\cite{Loth_Baumann_science_2012} and the implementation of the NAND gate with two antiferromagnetic spin chains\cite{Khajetoorians_Wiebe_science_2011}  have been demonstrated.

Spin waves are relevant excitations  in  systems
that display a  ground state with  well defined atomic spin magnetic moments,  such as  ferromagnets,  antiferromagnetic N\'eel  states,  spin spiral states, and skyrmions.   Here we study the spin waves of finite size spin chains that present a classical spin spiral ground state. 
Spin waves have been studied in a variety of  finite size  systems, including spin chains with ferromagnetic and antiferromagnetic ground states, \cite{Wieser2008,Wieser2009} as well as in skyrmions.\cite{Batista2013}
Our work is motivated by the recent experimental observation  of a stabilized noncollinear chiral  ground states  in  chains of Fe pairs deposited  on Ir(001) \cite{wiesendanger} (see Fig. 1).   This system\cite{Hammer}  has attracted interest both because of the non-trivial interplay between structure and magnetic coupling,\cite{Mazzarello2009,Mokrousov} and because  a local perturbation in one side of the chain affects the spin state globally,  as a consequence of long range spin, \cite{Onoda} as in the case of antiferromagnetically coupled spin chains.\cite{Khajetoorians_Wiebe_science_2011}   The robustness of spin spiral states against formation of domain walls is also considered an advantage.\cite{wiesendanger}  

Our interest on the spin waves in this system is twofold. First,  spin  excitations of spin chains, including spin waves, 
 could be probed by means of inelastic electron tunneling spectroscopy (IETS), \cite{Hirjibehedin_Lutz_Science_2006,JFR09,Lorente3,Delgado13} which would provide an additional experimental characterization of the system, complementary to spin polarized magnetometry.\cite{RMP-Wies} Second, spin waves are a source of quantum noise that sets a limit to the capability of sending spin information along the chain.   

The rest of this paper is organized as follows. In section II we briefly discuss the fundamentals of the spin spiral state ground state and the Hamiltonian used to describe it. In section III we discuss the method to compute the spin wave excitations.  In section IV we present the results of our numerical calculations. In section V we summarize our main conclusions.  

\section{Spin Spiral Hamiltonian}

Short range isotropic Heisenberg exchange naturally yields collinear spin alignments, either ferro or antiferromagnetic. 
The competition with a spin coupling that promotes perpendicular alignment, such as the antisymmetric Dzyaloshinskii-Moriya (DM) interaction\cite{DM1,DM2} $E_{DM}=\sum_{i,j}\mathbf{D}_{i,j}\cdot(\mathbf{S}_{i}\times\mathbf{S}_{j})$,  naturally results in a non-collinear  spin alignment between  first neighbors in the plane normal to $\mathbf{D}_{i,j}$. In one dimensional systems the DM term  leads to a spin spiral states and in two dimensions promotes the formation of skyrmions.\cite{Fert2013}   For the spin chains considered here, the vector $\mathbf{D}_{i,j}=(0,D,0)$  is the same for all couplings and lies along the $\hat{y}$ direction, perpendicular to the chain axis   $\hat{z}$ (see Fig. 1).  In these situations,   any global rotation  of the spin spiral in the (xz) plane   would result in a state with the same energy.  This large degeneracy is broken by the presence of  single ion uniaxial anisotropy term that  results in a preferred axis so that there are only two classical ground states.   In addition, the uniaxial anisotropy term distorts the spiral, preventing a uniform rotation angle along the chain. 
Finally, the  application of a magnetic field $\mathbf{B}$  along the $\hat{x}$ direction (perpendicular both to the chain axis and to $\vec{D}$)
 can further break the symmetry, resulting in a unique ground state.   These four terms are included in  the  Hamiltonian: 
\begin{eqnarray}
H=-\sum_{<i,j>}J_{i,j}\mathbf{S}_{i}\cdot \mathbf{S}_{j}+\sum_{<i,j>}\mathbf{D}_{i,j}\cdot(\mathbf{S}_{i}\times \mathbf{S}_{j})\nonumber
\\
+g\mu_B \mathbf{B}\cdot\sum_{i}\mathbf{S}_{i}-K\sum_{i}(S_{i}^{x})^{2},
\label{Hamiltonian1}
\end{eqnarray} 

We study  two types of chains. First  we consider a toy model of a mono strand  chain, with first neighbor couplings only,  $S=2$ and $D=J=1$ meV and $K=2$ meV. Then we move to a more realistic description of the diatomic Fe chains,\cite{wiesendanger} that includes couplings up to sixth neighbors, obtained from DFT calculations.    In both cases the classical ground state is calculated by minimizing the energy as a function of the magnetic configuration, defined by the orientation of the magnetic moments $\vec{S}_i$, that are treated as classical vectors whose lengths remain fixed.  The solutions are represented in the Figs. 1 and 2.

\begin{figure}
   \centering
\includegraphics[width=0.5\textwidth,angle=0]{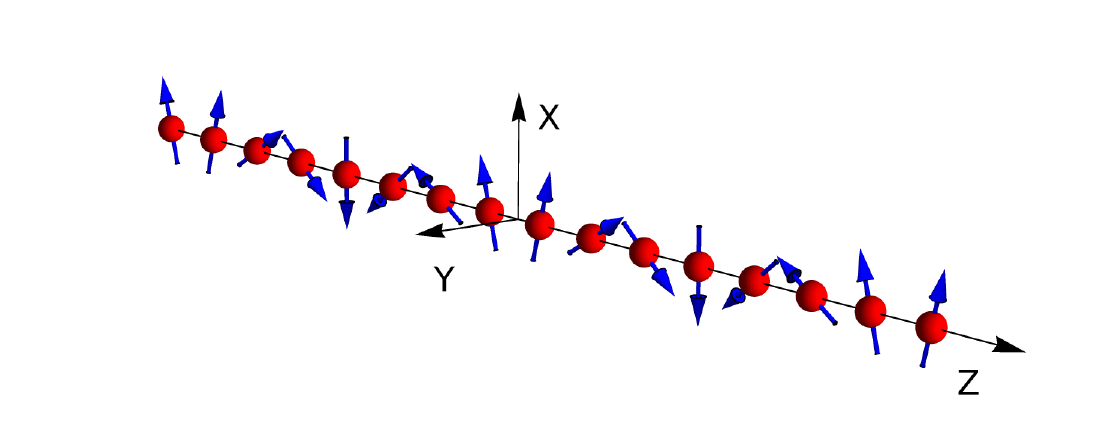}
\includegraphics[width=0.44\textwidth]{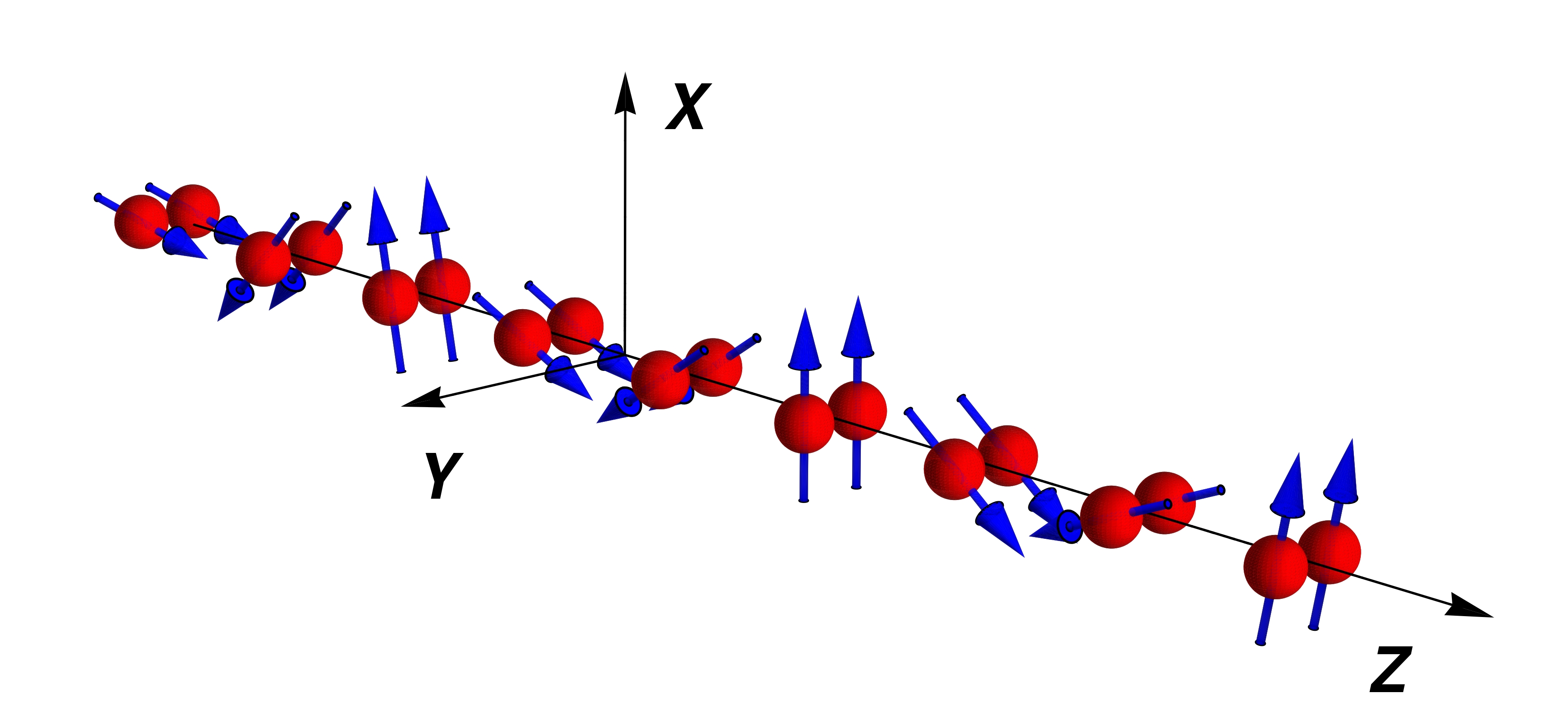}
\caption{ Schematic classical ground-state  for the mono strand (top) and the diatomic (bottom) chains. 
 The parameters used in the calculation are given in the text. In both cases a magnetic field of 2T along the $(1,0,0)$ direction is applied. }
\label{GS_scheme}
\end{figure}

\section{Calculation of spin waves}

The exact numerical diagonalization of the Hamiltonian (\ref{Hamiltonian1}), that would yield the spin excitations,  it is only possible in systems with a small number of atoms.  Therefore, we use the spin wave approximation. 
 The calculation of the spin wave spectrum of the  finite size chains is based on the representation of the spin operators
 in terms of Holstein Primakoff (HP) bosons:\cite{Auerbach,HP}
\begin{eqnarray}
\mathbf{S}_i\cdot\vec{\Omega}_i\!&=&\!S-a_{i}^{\dagger}a_{i},
\label{HPtransf}
\\
S_{i}^{+}\!&=&\!\sqrt{2S-n_{i}}\ a_{i}, \\
S_{i}^{-}\!&=&\!a^{\dagger}_{i}\sqrt{2S-n_{i}}
\end{eqnarray}

where $\vec{\Omega}_{i}$ is the spin direction of the classical ground state on the position $i$ and $a_i^{\dagger}$ is a bosonic creation operator and $n_i=a^{\dagger}_ia_i$ is the boson number operator.  The operator $n_i$   measures  the deviation of the system from the classical ground state.

The essence of the spin wave calculation is to  replace the spin operators in Eq. (\ref{Hamiltonian1}) by the HP representation and 
the truncation up to quadratic order in the bosonic operators.  Terms linear in the bosonic operators vanish when the expansion is done around the correct classical ground state.   This approach has been widely used in the calculations of spin waves for ferromagnetic and antiferromagnetic ground states.\cite{Yosida,Auerbach}
A generalized technique for HP approach in non-collinear systems has been developed. \cite{noncollinear2}  After a lengthy calculation detailed in the appendix, 
we obtain the following spin wave Hamiltonian: 
\begin{eqnarray}
H_{SW}=\sum_i m_{i}a_{i}^{\dagger}a_{i}+ \mu_{i}a_{i}^{\dagger}a_{i}^{\dagger} + {\rm h. c.}  \nonumber\\
+\sum_{<i,j>}t_{i,j}a_{i}^{\dagger}a_{j}+\tau_{i,j}a_{i}^{\dagger}a_{j}^{\dagger}+{\rm h.c.}
\label{Hamiltonian2}
\end{eqnarray}        

The specific values of elements $t_{i,j}$, $\tau_{i,j}$, $m_{i}$ and $\mu_{i}$ in the spin-wave Hamiltonian depend on the parameters of the interactions explicitly and implicitly by the classical ground state in the corresponding sites. These  are calculated in the appendix. In the  case of a collinear ferromagnetic ground state,  the anomalous terms that do not conserve the boson number vanish: $\tau_{i,j}=\mu_{i}=0$ .  In general,  in a non-collinear  ground state, the anomalous terms are different from zero and the magnon number is no longer a conserved quantity. In these cases a Bogoliubov approach is needed in order to diagonalize the Hamiltonian. We undertake such task  following the algorithm described in Ref. \onlinecite{Colpa}.  By so doing, we can  write the Hamiltonian (\ref{Hamiltonian2}) in the form: 
\begin{equation}
H_{SW}=\left[\!\!\begin{array}{cc}\chi^{\dagger}&\!\widetilde{\chi}\end{array}\!\!\right]\bar{\mathbf{H}}\left[\!\begin{array}{cc}\chi \\
\widetilde{\chi}^{\dagger}\end{array}\!\right]-\frac{1}{2}Tr[\bar{\mathbf{H}}]
\label{eq4}
\end{equation}
where $\chi^{\dagger}=[\!\!\begin{array}{cccc}a_{1}^{\dagger}\!\quad\!\! a_{2}^{\dagger}\dots a_{N}^{\dagger}\end{array}\!\!]$. The Hermitian matrix $\bar{\mathbf{H}}$ is a $2N\times 2N$ Bogoliubov-de Gennes Hamiltonian that must be diagonalized in terms of paraunitary transformation matrix\cite{Colpa} $\mathcal{T}$. This yields the diagonal form:
\begin{eqnarray}
H_{SW}=\left[\!\!\begin{array}{cc}\zeta^{\dagger}&\!\widetilde{\zeta}\end{array}\!\!\right]\frac{1}{2}\left[\begin{array}{cc}\bar{\omega} &  0_{N\times N}\\0_{N\times N}  & \bar{\omega}\end{array}\right]\left[\begin{array}{cc}\zeta \\
\widetilde{\zeta}^{\dagger}\end{array}\!\right]
\end{eqnarray} 
where $\bar{\omega}$ is a $N\times N$ diagonal matrix with   the spin wave spectrum $\omega_{\eta}$ 
and $\zeta^{\dagger}=[\alpha_{1}^{\dagger}\ \alpha_{2}^{\dagger}\cdots\alpha_{N}^{\dagger}]$ are the operators that create
the corresponding spin wave excitations. Their relation to the original HP bosons is: 
\begin{equation}
\left[\!\!\begin{array}{cc}\zeta \\
\widetilde{\zeta}^{\dagger}\end{array}\!\!\right]=\mathcal{T}\left[\!\!\begin{array}{cc}\chi \\
\widetilde{\chi}^{\dagger}\end{array}\!\!\right]\label{transformation}
\end{equation}
Thus,  the  ground state is defined by: $\alpha_{j}|GS\rangle=0$ for all $j$ and, in general, 
 is not the same than the classical ground state. For a given magnonic state $|\psi_{\eta}\rangle\equiv \alpha^{\dagger}_{\eta}|GS\rangle$,  the deviation from the classical ground state  at site $i$ is given by 
  $\rho_{i,\eta}=\langle \psi_{\eta}|a^{\dagger}_{i}a_{i}|\psi_{\eta}\rangle$.
Importantly, this quantity is non-zero even in the ground state, $\rho_{i,GS}=\langle GS|a^{\dagger}_{i}a_{i}|GS\rangle$, reflecting the zero-point quantum fluctuations that are a consequence of a noncollinear classical ground state and lead to a reduction of the magnetization along the classical direction (see eq. \ref{HPtransf}).

\section{Results}
We now apply the formalism of the previous section to compute the spin waves of finite size chains with spin spiral ground states.
This method has been applied to infinite crystals,  providing the spin wave dispersion $\omega(q)$ associated to spin spirals.\cite{Zheludev99}
%
%
\subsection{First nearest neighbor interaction monostrand chain.}
We address first the case of a simple chain with first nearest neighbors exchange and DM interactions.
In spite of its simplicity, we shall see that this simple model captures the essence of the physical behavior of the spin waves in the more realistic case described in the next subsection.
The first step is to calculate the classical ground state.  For a given choice of Hamiltonian parameters, the ground state is found either by a self-consistent minimization procedure 
or by classical Monte Carlo.   The  ground state of the mono strand chain is shown Fig. 1a and also in Fig. 2,   for uniaxial anisotropy $K=2$ meV, $S=2$, and  $J=D=1$ meV. 
It is apparent that, because of the single ion anisotropy term,  the spin spiral is distorted.  The choice of $\vec{D}=(0,D,0)$ yields a spin spiral in the $xz$ plane.  The period of the spiral is approximately 7 atoms, slightly shorter than the result obtained from the case without anisotropy ($2 \pi/\arctan(\frac{D}{J})$).

\begin{figure}
   \centering
\includegraphics[width=0.45\textwidth]{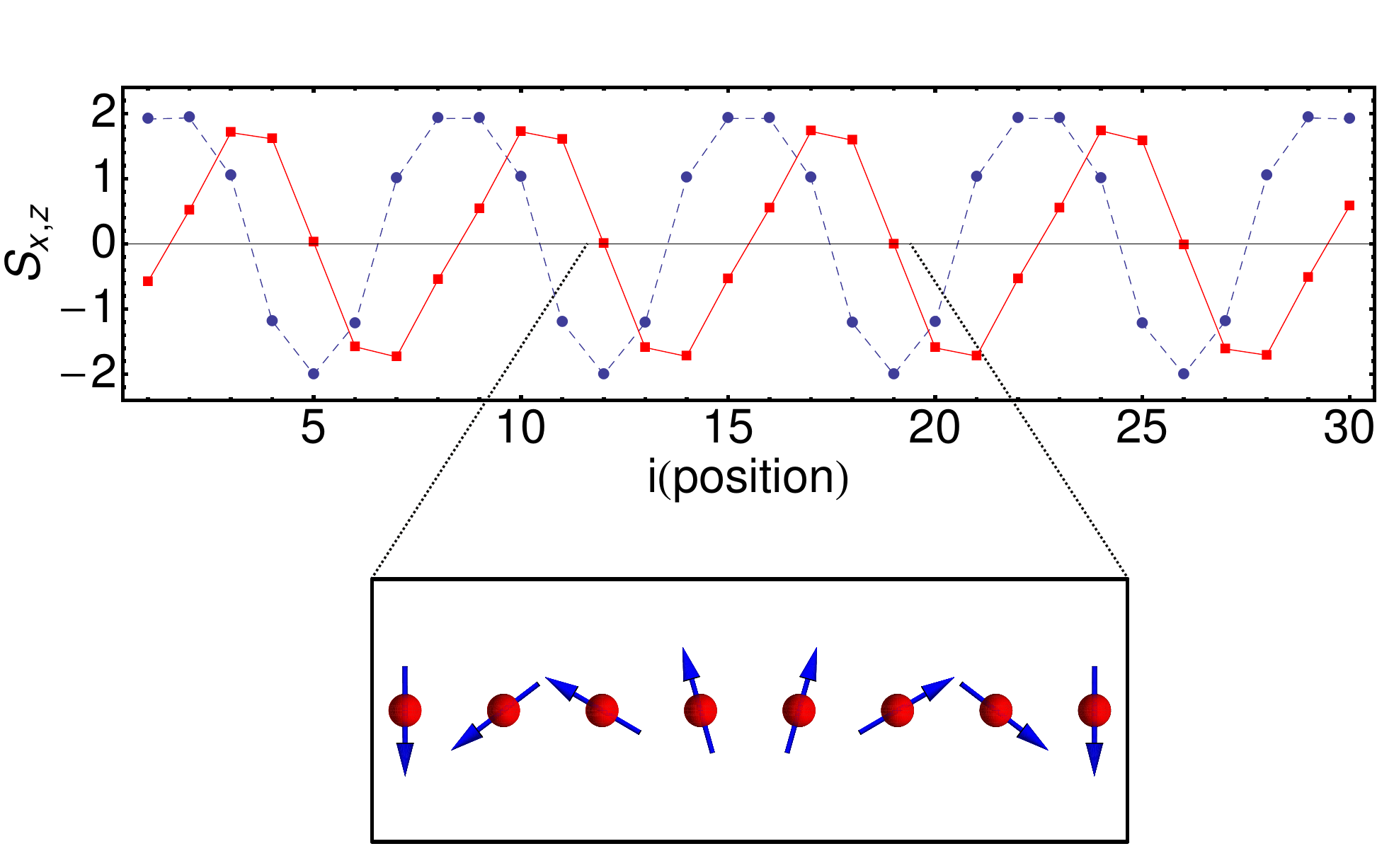}
\caption{Projections of the classical ground state $\vec{\Omega}_i$ over the $\hat{x}$ (dashed blue) and the $\hat{z}$ (solid red) directions for the monostrand chain with first nearest neighbor interactions. In this case we take $B=1T$, $J=D=1$ meV.}
\label{fig: Sxz_1_1}
\end{figure}
 Once  the classical ground state is determined, the problem is reduced to diagonalize the Hamiltonian in Eq. (\ref{Hamiltonian2}). This is achieved using the Bogoliubov-de Gennes prescription described in the previous section.
We focus on the  spin wave spectrum of finite size chains with $N=30$ sites.  In figure 3a, we show the evolution of the five
lowest energy modes as a function of the applied field $B$, applied along the easy axis ($\hat{x}$).  
 The abrupt change in the spectrum at fields near  $2$ T  corresponds to a drastic modification of the ground state from helical to ferromagnetic order. This  can be seen in Fig.  3(b) where we show the dependence of the net magnetization  along the $x$ axis for the classical ground state as function of field.  The jump in the spin wave spectrum takes place at the same field than the jump in the magnetization.  %

In Figs. 3c and 3d we show the expectation value of the HP boson occupation number $\langle a^{\dagger}_ia_i\rangle$ calculated within the spin wave vacuum (Fig 3c) and the lowest energy  spin wave state (Fig. 3d) as a function of both the applied field (vertical axis) and chain site (horizontal axis).  
The first thing to notice is that, in the spin spiral state, the quantum spin fluctuations are present even in the ground state.  These fluctuations disappear  in the FM ground state.   The ground state fluctuations present an oscillation across the chain, commensurate with the spin spiral.   The character of the first excited spin wave also changes from a edge mode in the spin spiral, to a extended state with 
a magnon density proportional to $\sin[\pi \frac{i}{N}]$.
 

\begin{figure}
   \centering
\includegraphics[width=0.22\textwidth]{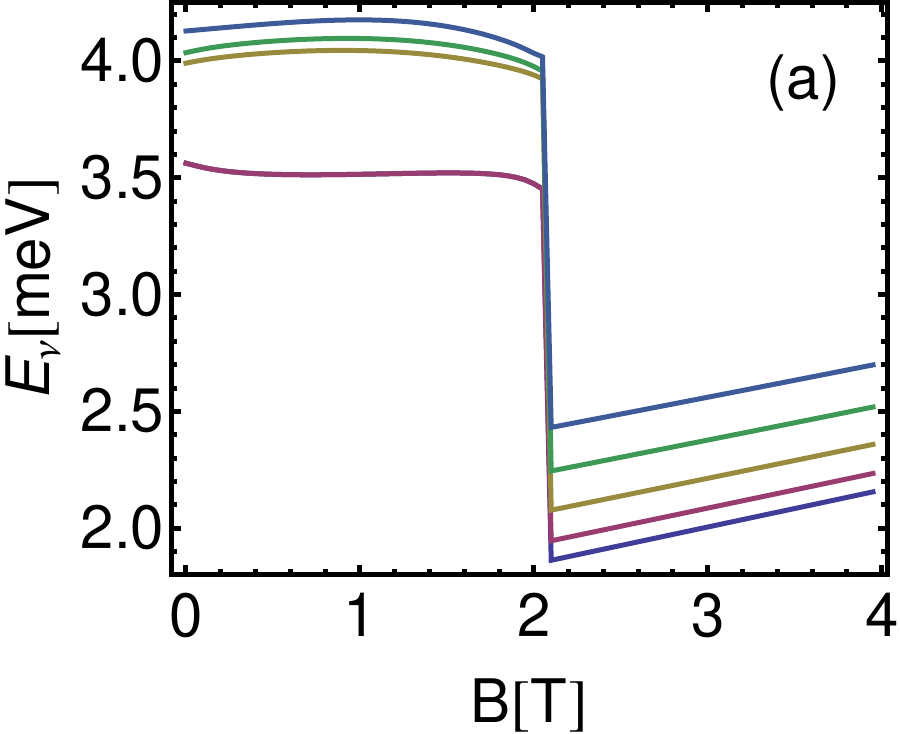}
\ \includegraphics[width=0.22\textwidth]{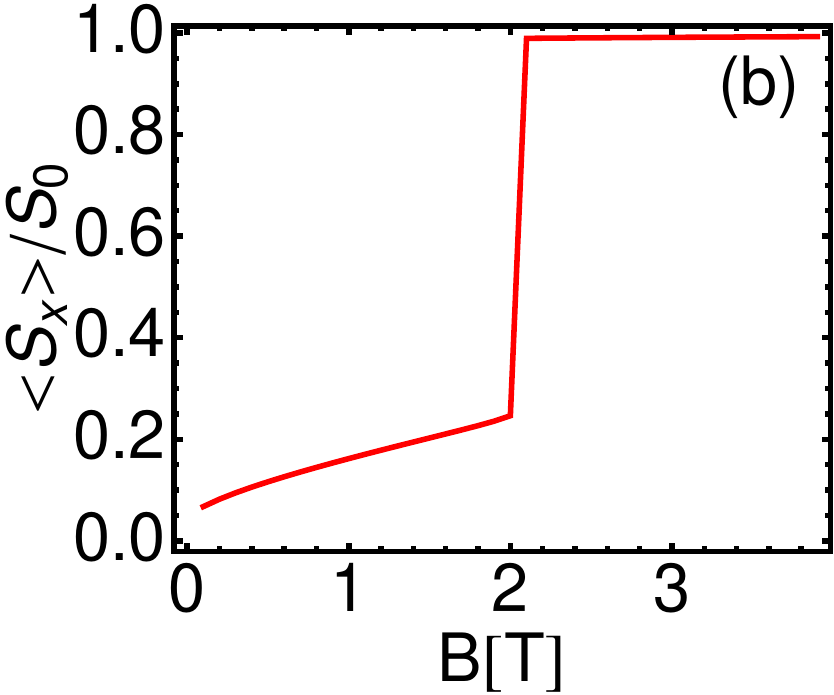}\\
\ \includegraphics[width=0.22\textwidth]{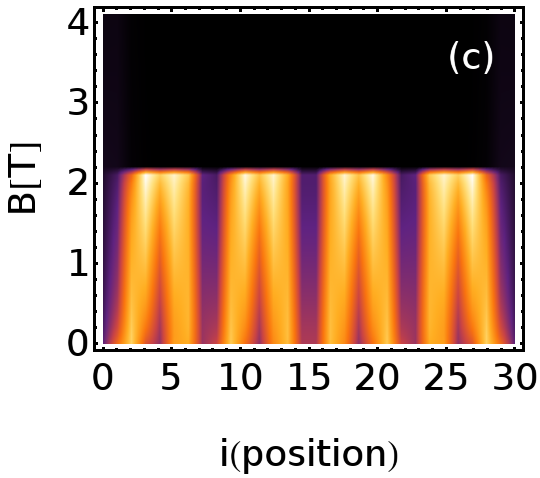}
\ \ \includegraphics[width=0.22\textwidth]{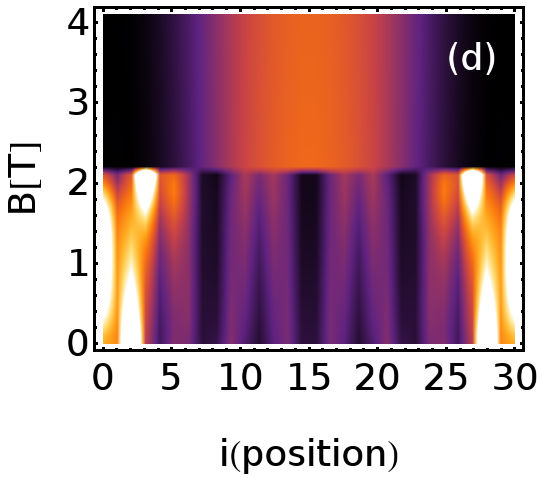}
\caption{Magnonic excitation analysis for the monostrand nearest neighbor interaction chain as function of magnetic fields along $x$ axis. a) Energies for the first five excited states a), dependence of the net spin along the $x$ axis for the classical ground state b), and magnonic occupation on the ground state c) and the firs excited state d).
In these plots, black stands for null  $\langle a^{\dagger}_ia_i\rangle$  and white for maximal fluctuation.
}
\label{fig: nearest neighbor}
\end{figure}

\subsection{Real Fe bi-atomic chain on Ir(001).}

We now compute the spin wave spectrum of  the bi-atomic Fe chains, described with a realistic spin Hamiltonian, 
obtained by fitting DFT calculations,  further validated by comparison with the experimental observations.\cite{wiesendanger}
The exchange and DM  parameters so obtained  include interactions up to  the six nearest neighbors (see table in the Appendix). 
 Interestingly, the results for the realistic model are qualitatively consistent with the findings of the simpler toy model of the previous subsection. We consider a chain with $N=30$  Fe dimers. 
  The ferromagnetic coupling inside a given Fe dimer is denoted by $J_p= 160$ meV,\cite{wiesendanger}  and is the dominant energy scale in the problem. As a result, the spins in the dimer are parallel.   The spin order along the chain is given by a spin spiral with period 3, as shown in Fig. 1b.  
  
  In analogy with the results of the previous subsection, in Fig. 4a  we also show the evolution of the 5 lowest energy spin waves as a function of the applied magnetic field along the $\hat{x}$ direction.   
  These modes evolve smoothly up to a critical field   ($B\sim 28$T) where an abrupt change takes place, corresponding to a phase transition from the helical state to a collinear ferromagnetic ground state.   This phase transition is also revealed  in Fig. 4b, where we show the total magnetization along the field direction as a function of the field strength. Our calculations show  an abrupt change of behavior at the critical field.

In analogy with the results of previous subsection,  in Figs. 4c and 4d we show the expectation value of the HP boson occupation number $\langle a^{\dagger}_ia_i\rangle$ calculated within the spin wave vacuum (Fig 4c) and the lowest energy  spin wave state (Fig. 4d) as a function of both the applied field and chain site.  In this case is also true that 
quantum spin fluctuations are present even in the ground state and  disappear  in the FM ground state.  The main differences between the two cases are the following. First, the modulation in the intensity of the quantum spin fluctuations across the chain  have a different period that corresponds to the different wavelength of the spin spiral. 
Second, the quantum spin fluctuations of the spin wave state in the FM state (at high field) have a fine structure, compared with their  mono strand analogue,  that arises from the coupling beyond first neighbors.

\begin{figure}
   \centering
\ \ \includegraphics[width=0.22\textwidth]{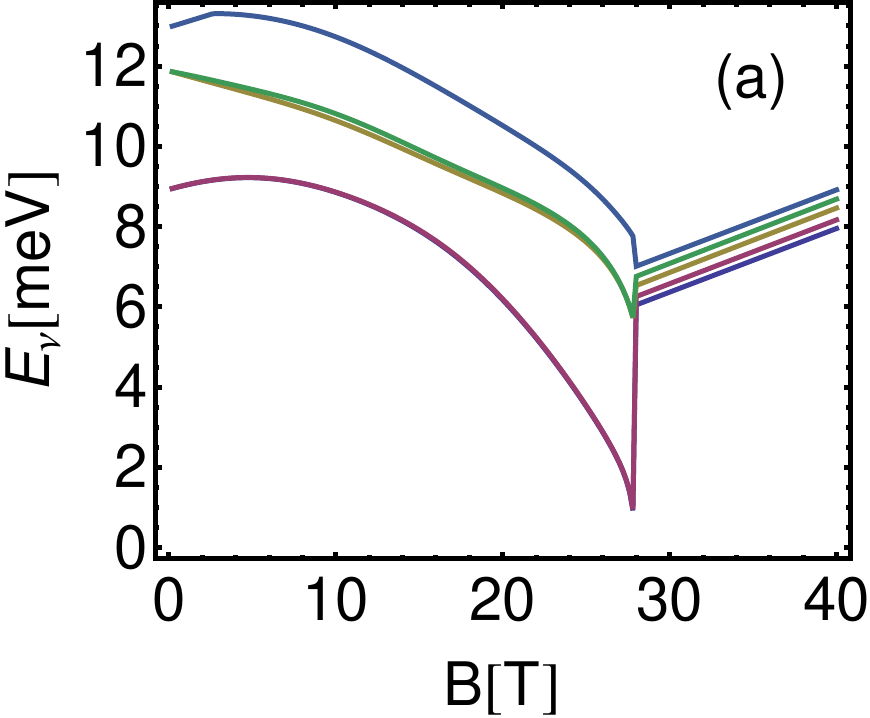}
\ \ \ \includegraphics[width=0.23\textwidth]{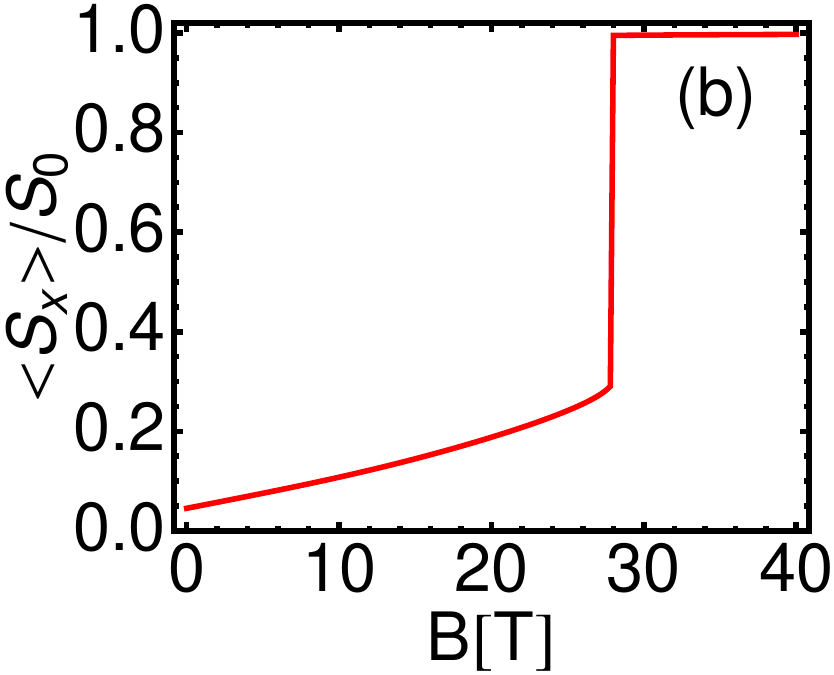}\\
\includegraphics[width=0.235\textwidth]{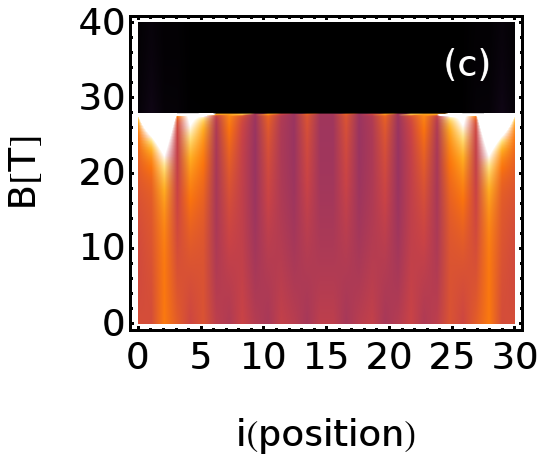}
\includegraphics[width=0.235\textwidth]{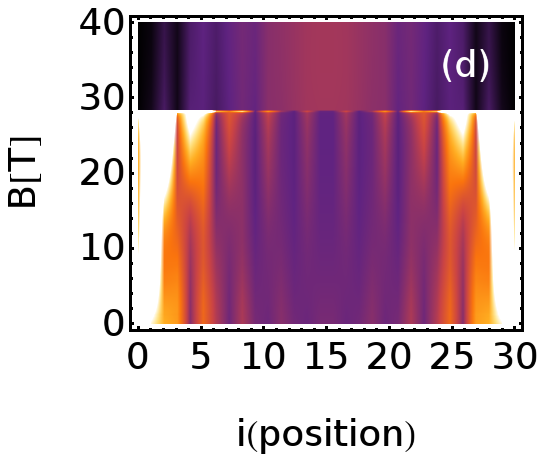}
\caption{Magnonic excitation properties for a 30 sites bi-atomic Fe chain on Ir(001) as a function of magnetic field along $x$ axis. Energies for the first five excited states a), dependence of the net spin along the $x$ axis for the classical ground b), and magnonic occupation on the ground state c) and the first excited state d).
In these plots, black stands for null  $\langle a^{\dagger}_ia_i\rangle$  and white for maximal fluctuation.
}
\label{fig: wies chain}
\end{figure}

We now discuss how the spin wave spectrum depends on the other parameter that can be controlled experimentally, namely, the number of dimers in the diatomic chain, $N$.  In fig. 5 we show the evolution of the 6 lowest spin wave energies $E_{\nu}$  as a function of $N$ for the spin spiral state (left panel) and the ferromagnetic state (right panel).   It is apparent that, for the spin spiral,  the $E_{\nu}$ present oscillations commensurate with the period of the spin spiral (3 dimers).  The plot of Fig. 4d, together with the   evolution of these first two  spin wave energies as a function of $N$,  suggest that they are edge modes. Their splitting at small $N$ arises from the hybridization of the two edge modes.  Therefore, a STM  could excite more easily the excitation of this mode when acting upon the edge atoms (something similar has been reported in reference \onlinecite{Delgado13}).   Future work will determine if this could result in an effective way to manipulate the spiral. 
In contrast, as soon  as $N$ is significantly  larger than the range the exchange interactions, the evolution of the $E_{\nu}$ in the ferromagnetic case displays a monotonic decrease as a function of $N$, consistent with the picture of confined bulk modes.

\begin{figure}
   \centering
   \includegraphics[width=0.23\textwidth]{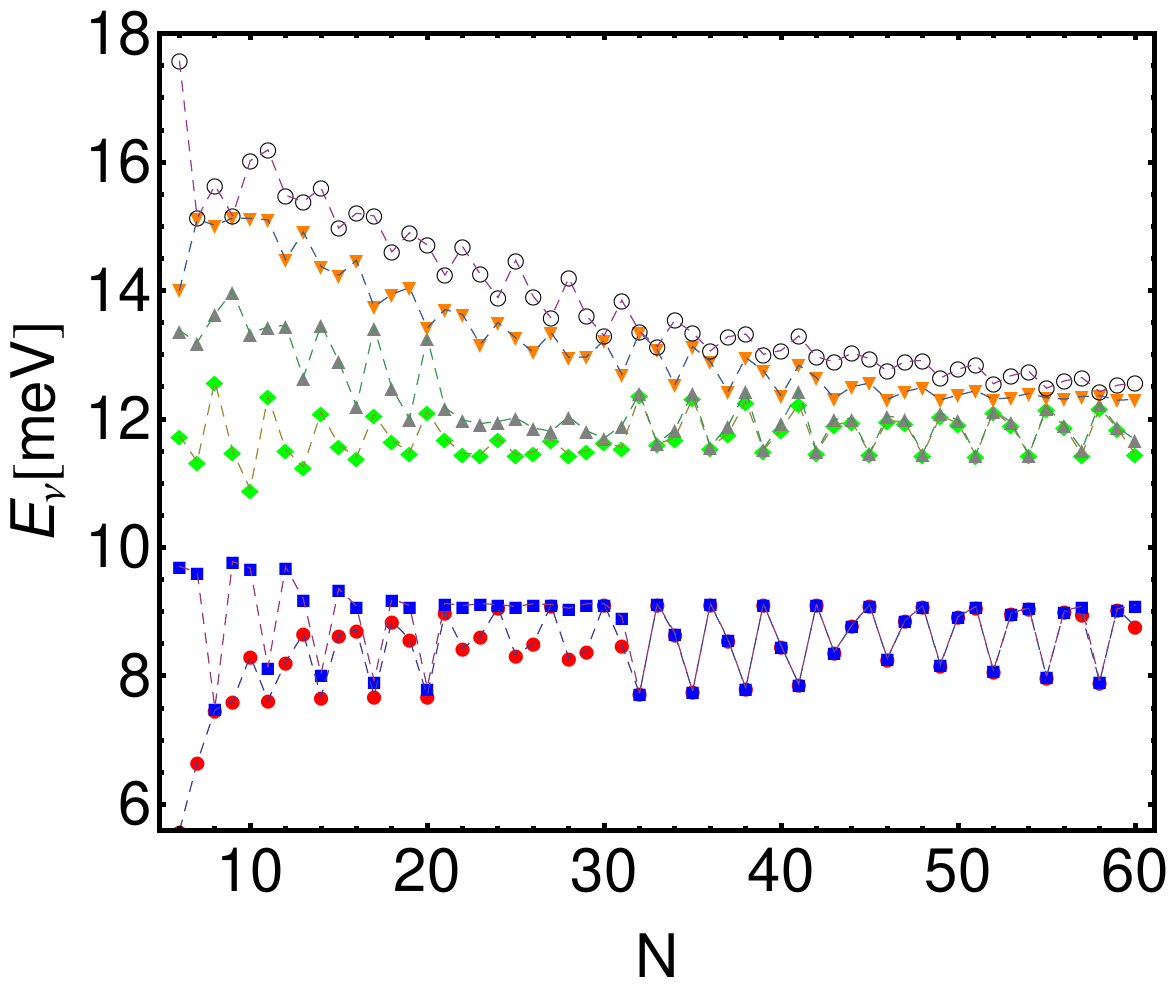} 
   \includegraphics[width=0.23\textwidth]{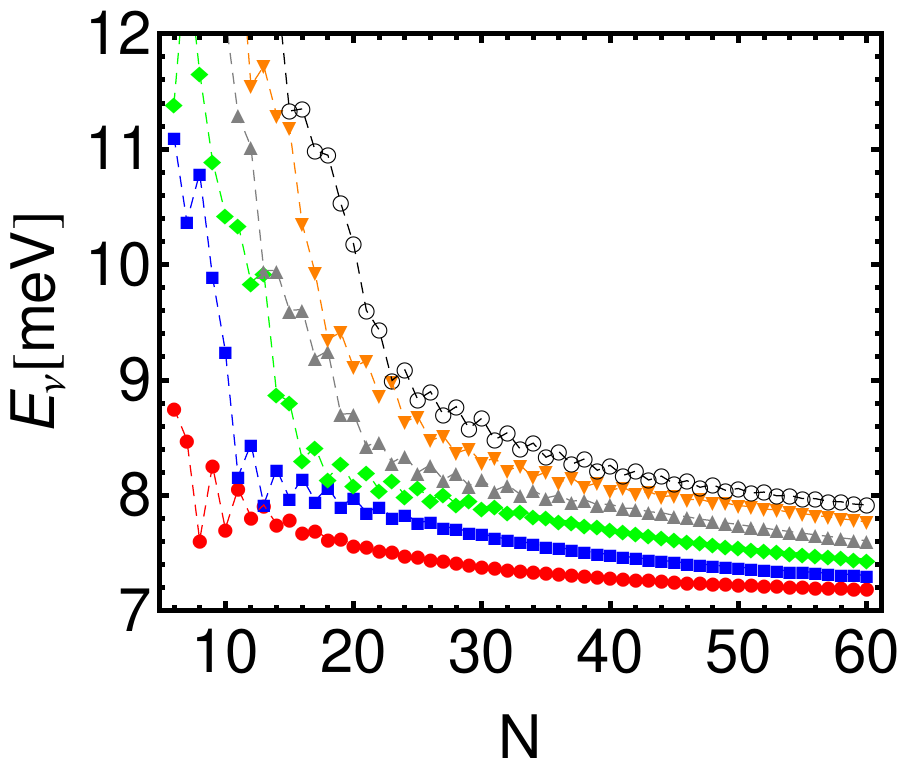} 
   \caption{Six lowest lying excitation energies as function of the chain's length for the spin spiral (left panel) and the ferromagnetic ground state (right panel). The calculation is 2T and 32T respectively. }
   
   \label{fig: spectrum_N}
\end{figure}

\section{Summary and conclusions}
We have studied the effect of spin wave excitations on the magnetic properties of finite size spin spirals, as those observed in recent experiments.\cite{wiesendanger}  
%
%
We have considered both a simple model with one atom per unit cell and first neighbors interactions as well as a more realistic\cite{wiesendanger} model with up to sixth neighbor couplings and two atoms in the unit cell.  In both cases we find three  interesting results. First, application of a magnetic field results in a phase transition from a spin spiral state at low field to a  ferromagnetic  state above a critical field. Second, the 
spin spiral ground state has zero point fluctuations that induce a reduction of the magnetization.  These zero point fluctuations are absent in the ferromagnetic state.  Third, the two lowest energy spin waves of the spin spiral are edge modes, in contrast to the bulk character of  the spin waves in the ferromagnetic case. 
 Our findings could be verified by means of inelastic electron tunneling spectroscopy. The existence of edge modes might provide a tool to manipulate the spin spiral by means of selective excitation of edge atoms with STM.

\section*{Acknowledgement}
We acknowledge F. Delgado for fruitful discussions and careful  reading of the manuscript. 
The authors would like to thank funding from grants Fondecyt 1110271, ICM P10-061-F by Fondo de Innovaci\'on para la Competitividad-MINECON and Anillo ACT 1117. ASN also acknowledges support from Financiamiento Basal para Centros Cient\'ificos y Tecnol\'ogicos de Excelencia, under Project No. FB 0807(Chile).  ARM,  MJS and ASN  acknowledge hospitality of INL. 

\appendix
\section{ Holstein-Primakoff Hamiltonian in noncollinear ground state.}

The HP  representation of the  spin operator discriminates one direction (see for instance Eq. 2) which is normally given by the magnetization of the classical ground state. Here we describe the technical details related to the use of HP bosons to compute the effective Hamiltonian in the case of non-collinear classical ground states.   For that matter, it is convenient to  define a rotated local coordinate system as follows:
\begin{eqnarray*}
\hat{e}_{i}^{1}&=&\cos\theta_{i}\ \cos\phi_{i}\ \hat{x}+\cos\theta_{i}\ \sin\phi_{i}\ \hat{y}-\sin\theta_{i}\ \hat{z}\\
\hat{e}_{i}^{2}&=&\hat{e}^3_{i}\times\hat{e}_{i}^{1} \\
\vec{\Omega}_i=\hat{e}^{3}_i&=&\sin\theta_{i}\ \cos\phi_{i}\ \hat{x}+\sin\theta_{i}\ \sin\phi_{i}\ \hat{y}+\cos\theta_{i}\ \hat{z}
\end{eqnarray*}

or, in a more compact form:
\begin{eqnarray}
\hat{e}_{\alpha}^{i}=(R_{\alpha,\beta}^{i})^{-1}\ \hat{r}^{\beta}
\end{eqnarray}
where the angles $\theta_{i}$ and $\phi_{i}$ characterize the spin direction on the classical ground state in the site $i$ and $\hat{r}^{\beta}$ are the cartesian axis. 
In this framework, the Hamiltonian (\ref{Hamiltonian1}) is 
expressed as:
\begin{eqnarray}
H_{ex}&=&\sum_{i,j}J_{i,j}\vec{S}_{i}\cdot \vec{S}_{j}\nonumber
\\&=&\sum_{i,j}J_{i,j}(\vec{S}_{i}\cdot \hat{e}_{\alpha}^{i}) (\vec{S}_{j}\cdot \hat{e}_{\beta}^{j}) R_{\alpha,\gamma}^{i} R_{\beta,\gamma}^{j}
\label{Hparameters1}
\\H_{DM}&=&\sum_{i,j}\vec{D}_{i,j}\cdot\vec{S}_{i}\times\vec{S}_{j}\nonumber
\\&=&\sum_{i,j}(\vec{S}_{i}\cdot \hat{e}_{\alpha}^{i}) (\vec{S}_{j}\cdot \hat{e}_{\beta}^{j}) R_{\alpha,\gamma}^{i}
\bar{D}_{i,j}^{\gamma,\eta} R_{\eta,\beta}^{j}
\end{eqnarray}
where we have defined $\bar{D}_{i,j}^{\gamma,\eta}\equiv D_{i,j}^a\epsilon_{\gamma,\eta,a}$ and $\epsilon_{\gamma,\eta,a}$ is the Levi-Civita symbol   and  a sum over repeated indexes is understood.
\begin{eqnarray}
H_{A}&=&-\sum_{i,\alpha}K_{\alpha}(S_{i}^{\alpha})^{2}\nonumber
\\&=&-\sum_{i}(\vec{S}_{i}\cdot \hat{e}_{\alpha}^{i}) (\vec{S}_{i}\cdot \hat{e}_{\beta}^{i}) R_{\alpha,\gamma}^{i}\bar{K}_{i}^{\gamma,\eta} R_{\eta,\beta}^{i}
\\H_{Zee}&=&\mu_{s}\sum_{i,\alpha}\vec{B}_{i}\cdot \vec{S}_{i}
=\mu_{s}\sum_{i}(\vec{S}_{i}\cdot \hat{e}_{\alpha}^{i}) R_{\alpha,\gamma}^{i}B_{i}^{\gamma}
\label{Hparameters2}
\end{eqnarray}

In the specific case considered here, where  $\vec{D}_{i,j}\parallel \hat{y}$ and $\hat{x}$ direction as anisotropic easy axis, we have:
\begin{eqnarray}
\bar{D}_{i,j}^{\gamma,\eta}&=&D_{i,j}(\delta_{\gamma,3}\delta_{\eta,1}-\delta_{\gamma,1}\delta_{\nu,3})
\\
\bar{K}_{i}^{\gamma,\eta}&=&K_{i}\ \delta_{\gamma,1}\ \delta_{\eta,1}
\label{DK}
\end{eqnarray}

The derivation of the  HP  Hamiltonian of Eq. (\ref{Hamiltonian2}). 
starts with the combined use of Eq. (\ref{HPtransf}) and the  expressions:
\begin{eqnarray}
S_{i}^{\pm}&=&\vec{S_{i}}\cdot \hat{e}_{1}\pm i\vec{S_{i}}\cdot \hat{e}_{2}\\
S-n_{i}&=&\vec{S_{i}}\cdot \vec{\Omega}_i
\end{eqnarray}
By inserting this in Eq. (\ref{Hparameters1}\ref{Hparameters2}), and keeping up to second order terms in the bosonic operators $a^{\dagger},a$,  we  are able to write the effective spin wave Hamiltonian  in the form of Eq. (\ref{eq4}), with:
\begin{equation}
\bar{\mathbf{H}}=\left[\begin{array}{cc}\mathbf{A} & \mathbf{B} \\
\mathbf{B^{*}} & \mathbf{A^{*}}
\end{array}\right]
\end{equation}
where $\mathbf{A}$ and $\mathbf{B}$ are $2N\!\times\!2N$ hermitic and symetric matrix, respectively. In the specific case of the diatomic chain, the elements of the matrices $A$ and $B$  read:
\begin{displaymath}
A_{i,j}=\delta_{i,j}\left[\begin{array}{cc} m_{i} & t_{p} \\
t_{p} & m_{i}
\end{array}\right]+(1-\delta_{i,j})\left[\begin{array}{cc}t_{i,j} & 0 \\
0 & t_{i,j}
\end{array}\right]
\end{displaymath}
\begin{displaymath}
B_{i,j}=\delta_{i,j}\left[\begin{array}{cc} \mu_{i} & 0 \\
0 & \mu_{i}
\end{array}\right]+(1-\delta_{i,j})\left[\begin{array}{cc}\tau_{i,j} & 0 \\
0 & \tau_{i,j}
\end{array}\right]
\end{displaymath}
where
\begin{eqnarray}
m_{i}&=&\frac{S}{2}\sum_{j=-6}^{6}(J_{i,i+j}\cos[\theta_{i}-\theta_{i+j}]+ D_{i,i+j} \sin[\theta_{i}-\theta_{i+j}])\nonumber \\&+&
 \frac{1}{2}K (2S-1-(3S-2) \cos^{2}[\theta_{i}])+\frac{h}{2} \sin[\theta_{i}]\nonumber \\
 &+&\frac{S}{2}J_{p}
\end{eqnarray}
 where $J_{ij}$ and $D_{ij}$ stand for inter-dimer exchange and DM coupling, $K$ arises from the unaxial single atom anisotropy and $J_p$ stands for the intra-dimer ferromagnetic exchange.  The values for $J_{ij}$ and $D_{ij}$ are given in table I.\cite{wiesendanger}  The other terms in matrices $A$ and $B$ are: 
 \begin{eqnarray}
\mu_{i}&=&-\frac{S}{2}K\sqrt{1+\frac{1}{2S}}\cos^{2}\theta_{i} \nonumber\\
t_{i,j}&=&-\frac{1}{4}S(J_{i,j}(1+\cos[\theta_{i}-\theta_{j}])+D_{i,j}\sin{[\theta_{i}-\theta_{j}]}) \nonumber\\ 
\tau_{i,j}&=&\frac{1}{4}S(J_{i,j}(1-\cos[\theta_{i}-\theta_{j}])-D_{i,j}\sin{[\theta_{i}-\theta_{j}]})  \nonumber\\ 
t_{p}&=&-\frac{S}{2}J_{p}
\end{eqnarray}

\begin{table}
\begin{tabular}{||c|c|c|c|c|c|c||}
\hline
\hline
$|i-j|$ & 1 & 2 & 3 & 4 & 5 & 6 \\
\hline
$J_{i,j}$(meV) & 0.53 & 1.42 & -0.12 & -0.34 & -0.29 & 0.37 \\
\hline
$D_{i,j}$(meV) & 2.58 &  -2.77 &  -0.07 &  0.63 &  -0.36 &  -0.12 \\
\hline
\hline
\end{tabular}
\label{table1}
\caption{Exchange and DM constants extracted from fits to the DFT calculations \cite{wiesendanger}}
\end{table}


\begin{thebibliography}{50}


\bibitem{Hirjibehedin_Lutz_Science_2006}  C. F. Hirjibehedin {\em et al}, Science {\bf 312}, 1021 (2006).

\bibitem{Serrate_Ferriani_natnano_2010} D. Serrate {\em et al.},  Nature Nanotechnology {\bf 5},  350 (2010).

\bibitem{Khajetoorians_Wiebe_science_2011} A. Khajetoorians {\em et al.},  Science {\bf 332}, 1062 (2011).

\bibitem{wiesendanger} M. Menzel, Y. Mokrousov, R. Wieser, J. E. Bickel, E. Vedmedenko, S. Bl\"ugel, S. Heinze, K. von Bergmann, A. Kubetzka \& R. Wiesendanger, Phys. Rev. Lett. {\bf 108}, 197204 (2012).

\bibitem{Khajetoorians_Wiebe_natphys_2012} A. Khajetoorians {\em et al.}  Nat. Phys. {\bf 8}, 497 (2012).

\bibitem{Loth_Baumann_science_2012}
S. Loth {\em et al.}, Science {\bf 335}, 196 (2012).


\bibitem{Yosida}K. Yosida, \textit{Theory of Magnetism}, (Springer, Heidelberg 1996).

\bibitem{Auerbach}Assa Auerbach, \textit{Interacting Electrons and Quantum Magnetism}, (Springer, Ney York 1994).

\bibitem{JFR-NAT-MAT2013} J. Fern\'andez-Rossier,  Nat. Mat. {\bf 12}, 480 (2013).

\bibitem{Wieser2008} R. Wieser,  E. Y. Vedmedenko, and R. Wiesendanger, Phys. Rev. Lett. {\bf 101}, 177202 (2008).

\bibitem{Wieser2009} R. Wieser,  E. Y. Vedmedenko, and R. Wiesendanger, Phys. Rev. B {\bf 79}, 144412 (2009).

\bibitem{Batista2013} S.Z. Lin,  C. D. Batista, A. Saxena, arxiV:1309.5168.

\bibitem{Hammer}L. Hammer, W. Meier, A. Schmidt and K .Heinz, Phys. Rev. B {\bf 67}, 125422 (2003).

\bibitem{Mazzarello2009}  R. Mazzarello, E. Tosatti, Phys. Rev. B {\bf 79}, 134402 (2009).
\bibitem{Mokrousov}Y. Mokrousov, A. Thiess and S. Heinze, Phys. Rev. B {\bf 80}, 195420 (2009).

\bibitem{Onoda}  S. Onoda, Physics {\bf 5}, 53 (2012).



\bibitem{JFR09}J. Fern\'andez-Rossier, Phys. Rev. Lett. {\bf 102}, 256802 (2009).

\bibitem{Lorente3}J. P. Gauyacq and N. Lorente, Phys. Rev. B {\bf 83}, 035418 (2011).

\bibitem{Delgado13}  F. Delgado, C. D. Batista,  J. Fern\'andez-Rossier, Phys. Rev. Lett. {\bf 111}, 167201 (2013).

\bibitem{RMP-Wies}  
R. Wiesendanger, Rev. Mod. Phys. {\bf 81},  1495 (2009).

\bibitem{DM1} I. E. Dzyaloshinskii, J. Phys. Chem. Sol. {\bf 4}, 251 (1958).

\bibitem{DM2} T. Moriya, Phys. Rev. {\bf 120}, 91 (1960).


\bibitem{Fert2013} A. Fert,  V. Cross,  J. Sampaio, Nat. Nano. {\bf 8},  152 (2013).

\bibitem{HP}T. Holstein and H. Primakoff, Phys. Rev. {\bf 58}, 1098 (1940).


\bibitem{noncollinear2}J. T. Haraldsen and R. S. Fishman, J. Phys.: Condens. Matter {\bf 21} 216001 (2009).



\bibitem{Colpa} J. H. P. Colpa, Physica A {\bf 93}, 327-353 (1978). 


\bibitem{Zheludev99} A. Zheludev {\em et al.}, Phys. Rev. B {\bf 59}, 11432 (1999).










\end{thebibliography}
\end{document}